\documentstyle[12pt,preprint]{aastex}

\slugcomment{Accepted to ApJ Letters}

\lefthead{Potter, Martin, Cushing, Baudoz, Brandner, Guyon, Neuh\"{a}user}
\righthead{Hokupa'a-Gemini Discovery of Two Ultracool Companions}

\begin{document}

\title{Hokupa'a-Gemini Discovery of Two Ultracool Companions to the Young Star HD 130948\footnote{Based on observations obtained at the Gemini Observatory, which is operated by the Association of Universities for Research in Astronomy, Inc., under a cooperative agreement with the NSF on behalf of the Gemini partnership: the National Science Foundation (United States), the Particle Physics and Astronomy Research Council (United Kingdom), the National Research Council (Canada), CONICYT (Chile), the Australian Research Council (Australia), CNPq (Brazil) and CONICET (Argentina)}
}

\author{D. Potter, E. L. Mart\'\i n, M. C. Cushing, P. Baudoz, W. Brandner\altaffilmark{1},  O. Guyon, and R. Neuh\"{a}user\altaffilmark{2}}

\affil{Institute for Astronomy, University of Hawaii, 2680 Woodlawn Drive, Honolulu, HI 96822}
\altaffiltext{1}{European Southern Observatory, Garching GE}
\altaffiltext{2}{Max-Plank-Institut f\"{u}r extraterrestrische Physik, Garching GE}



\begin{abstract}

We report the discovery of two faint ultracool companions to the nearby (d$\sim$17.9~pc) 
young G2V star HD~130948 (HR~5534, HIP~72567) using the Hokupa'a adaptive 
optics instrument mounted on the Gemini North 8-meter telescope. Both objects  
have the same common proper motion as the primary star as seen over a 7 month baseline 
and have near-IR photometric colors that are consistent with an early-L classification. 
Near-IR spectra taken with the NIRSPEC AO instrument on the 
Keck II telescope reveal K I lines, FeH, and H$_2$O bandheads.  
Based on these spectra, we determine that both objects have spectral 
type dL2 with an uncertainty of 2 spectral subclasses.  
The position of the new companions on the 
H-R diagram in comparison with theoretical models is consistent with the young age of the primary star
($<$0.8~Gyr) estimated on the basis of X-ray activity, lithium abundance 
and fast rotation. HD~130948B and C likely constitute a pair of young contracting brown dwarfs 
with an orbital period of about 10~years, and will yield dynamical masses for L dwarfs in the near future. 

\end{abstract}




%

\section{Introduction}

The past decade of near-IR sky surveys and technological advances in high dynamic range imaging have 
found a large number ($\sim$100) of very low-mass (VLM), ultracool objects. This has brought about spectral classification 
schemes (Burgasser et al. 2001; Geballe et al. 2001; Kirkpatrick et al. 1999, 2000; Mart\'\i n et al. 1997, 1998, 1999b) 
attempting to organize and understand them in the same way as we understand  main sequence 
stars through the MK spectral classification scheme. However, the interpretation of physical parameters from the classification schemes is a more complicated 
exercise with ultracool objects as the lack of a sustained hydrogen burning core creates a degeneracy between mass and age as the luminosity 
fades in time. Also, the spectra of these objects are significantly affected by dust in their 
atmospheres (Allard et al. 2001; Basri et al. 2000; Schweitzer et al. 2001) possibly introducing a weather-like 
time variable phenomenon 
(Bailer-Jones \& Mundt 2001; Mart\'\i n et al. 2001; Nakajima et al. 2000). Just as stellar evolution theory 
was calibrated 
using the dynamical mass estimates of binary stars, it is important to check evolutionary tracks of VLM objects using low-luminosity binaries. 

In recent years, there have been surveys using the high resolution capabilities of the HST (Mart\'\i n et al. 1999a, 2000a; 
Reid et al. 2001) and of large ground based telescopes (Close et al. 2002; Koerner et al. 1999; Mart\'\i n et al. 2000b) 
to look for companions to the known VLM objects. 
One goal of these searches is to build a sample of VLM binary systems in which accurate dynamical masses can be obtained. A handful of brown dwarf binaries are known, but  
only Gl~569B (Lane et al. 2001, Kenworthy et al. 2001), 2MASSW~J0746425+200032 (Reid et al. 2001), and 2MASSJ 1426316+155701 (Close et al. 2002) have periods $\lesssim$10 years. 

In this paper we add to the growing list of VLM ultracool binary systems. In a companion search around nearby, 
young (less than 1 Gyr), solar-type stars selected from the sample of Gaidos et al. (2000), we found two companions 
next to the star HD~130948.  
\S2 overviews the observations, \S3 presents the photometric, astrometric, and spectroscopic results which 
confirm that the companions are truly associated with the primary star.\S4 discusses the placement of the objects on an HR-diagram compared with theoretical evolutionary models, and presents estimations of the age and mass of the companions.

\section{Observations and Data Reduction}

The two companions of HD~130948 were discovered using the Hokupa'a (Graves et al. 1999) curvature sensing Adaptive Optics (AO) 
system mounted on the Gemini North Telescope in the Wollaston prism mode (Potter et al. 2002, in preparation) 
on the night of 2001 February 24 (UT). Figure~1 displays the discovery image of the companion system which is separated by 2$\farcs$64$\pm$0$\farcs$01 and is 8 magnitudes fainter in the H-band relative to the primary star.. 
Before the photometric and astrometric analysis, all images were flat-fielded with the bad pixels filtered from the images. The field of view in the Wollaston prism mode is a rectangle that is 4$\arcsec$x20$\arcsec$. Three sets (20 exposures/set) of 20 second exposures were obtained to give a total exposure time of 1200 seconds. Each set of exposures was taken with different field orientations separated by 90$^\circ$ using Gemini's instrument rotator. No other stars were observed in the field.
 
To check for a common proper motion between the primary star and the new companions, and for possible orbital 
motion between the VLM binary pair, the objects were observed on four different occasions over a time baseline of 
204 days between 2001 February 24 (UT) and 2001 September 20 (UT). The proper motion of HD 130948 is well known (Perryman et al. 1997) to be 
148 mas/year, which equates to 4.3 pixels of relative movement between background stars and common proper motion objects on the 
Hokupa'a/QUIRC detector. We find there is no significant differential motion between the binary pair (B and C) and the primary star (A) 
within our astrometric accuracy of $\sim$5 mas. Therefore, the two objects are most likely a gravitationally bound pair at the same istance as the primary star (17.9 pc). 
The average astrometric result is a separation between HD~130948 B and C equal to $\rho$=0$\farcs$134$\pm$0$\farcs$005 and a position angle 
equal to PA=317$^o$$\pm$1$^o$. 

The $J$, $H$, $K$ photometry was obtained using the MKO near-IR photometric system based on the UKIRT faint standard star list (Hawarden et al. 2001). 
The photometry was obtained on 2001 April 19 (UT). The halo of the bright primary star presented an obstacle in obtaining accurate photometry of the companions. In order to subtract the profile of the halo, each image was differenced with a version of itself, rotated about the photocenter of the primary by 180$^o$. After the subtraction, the background immediately surrounding the companions was relatively flat which allowed the use of curve of growth aperature photometry on the combined light of both companions. The brightness ratio of the two companions was then estimated to be B/C=4/3 based on the profiles of the stars. The same parameters for the curve of growth method were used for the UKIRT faint standard, FS137. The absolute magnitudes based on a 17.9 pc distance and their errors are provided in Table 1. The measured colors of both objects are consistent with those of field L dwarfs (Leggett et al. 2001).

\begin{table}
\begin{center}
\caption{Photometry of HD~130948B and C}
\begin{tabular}{llll}
\hline
Component & M$_J$  & M$_H$ & M$_{K_s}$   \\
\hline
\hline
B	&  12.6   $\pm$0.2 & 11.9 $\pm$ 0.1  & 11.0 $\pm$ 0.1 \\	
C      	& 12.9   $\pm$ 0.2 &  12.3 $\pm$ 0.1 & 11.3 $\pm$ 0.1\\
\hline
\hline
\end{tabular}
\end{center}
\end{table}

Medium-resolution spectra (R=1500) from 1.15-1.35 $\mu$m of each component of the binary were obtained 
on the Keck~II 10-meter telescope using NIRSPEC (McLean et al. 2000) 
with AO on 2001 June 30 (UT). 
The AO correction was made using HD~130948 as the wavefront sensor guide star. The system delivered images with FWHM=0$\farcs$06 (3.4 pixels) 
in the $J$-band. 
HD~130948 B and C were clearly resolved, and the NIRSPEC 3 pixel wide slit was placed along the axis joining both objects. 
Three exposures of 300~s were obtained. An A0 V standard star, HD~131951, was observed inmediately after to correct for telluric 
absorption features. 

Data reduction was performed using IRAF tasks. The reduction procedure included sky subtraction, flat field division, 
aperture tracing, extraction of the one dimensional spectrum, wavelength calibration using a lamp spectrum, division 
by the normalized spectrum of the A0 star (after removing the P$_\beta$ absorption feature at 1.28~$\mu$m), and 
multiplication by a black body function for a temperature of 9500~K. 
Figure~2 displays the final NIRSPEC spectra of HD~130948B and C.  
We compared these NIRSPEC spectra with SpeX data of VLM dwarfs with spectral types in the range M8 to L5 
(Cushing et al. in preparation). 
The SpeX data have a resolution (R=2000) similar to the NIRSPEC data. 
We measured the strength of the absorption features indicated in Figure~3. 
Spectroscopic measurements are given in Table~2. 
Integration limits and equivalent widths (EW) for objects with spectral types 
in the range M7 to L5 are provided.   
The spectrum of HD~130948C is undistinguishable from that of HD~130948B, and thus 
we only give the EW values of the brighter component of the pair.

\begin{table}
\begin{center}
\caption{Spectroscopic measurements}
\begin{tabular}{llllll}
\hline
Name                    & SpT  & EW(KI) (\AA )    & EW(KI) (\AA )    & EW(FeH) (\AA )   & EW(H$_2$O) (\AA ) \\ 
$\Delta\lambda$ (nm)    &      & 1.1655-1.1715 & 1.1750-1.1810 & 1.1930-1.2080 & 1.3420-1.3600  \\
Pseudo-Continuum (nm)   &      & 1.1710-1.1750 & 1.1710-1.1750 & 1.1830-1.1930 & 1.2880-1.3020  \\
\hline
VB10                    & dM8  & 3.9           & 7.2           & 8.3           & 31.6           \\
DENIS-P~J104814-395606  & dM9  & 5.6           & 8.1           & 10.7          & 38.7           \\
2MASSW~J1439284+192915  & dL1  & 6.7           & 9.3           & 12.8          & 47.2           \\
Kelu~1                  & bdL2 & 6.0           & 8.2           & 14.1          & 44.4           \\
2MASSW~J1146345+223053  & bdL3 & 5.9           & 9.2           & 14.7          & 40.4           \\
2MASSW~J1507476-162738  & dL5  & 7.7           & 9.5           & 13.1          & 56.6           \\
HD~130948B              &      & 6.4           & 8.1           & 13.7          & 43.0           \\ 
\hline
\hline
\end{tabular}
\end{center}
\end{table}

On the basis of the measurements shown in Table~2, and the visual comparison with standard 
spectra, we estimated that both 
HD~130948B and C are cooler than M9 and warmer than L5. However, 
the relatively narrow spectral region covered with NIRSPEC does not allow us to distinguish between spectral 
subclasses in the range L0 to L4. 
Thus, we adopt a spectral type of dL2 with an uncertainty of 2 spectral subclasses for both 
HD~130948B and C\footnote{We adopt the Mart\'\i n et al. (1999) notation of dL2 for dwarfs of spectral class L2. 
We note that the spectral subclass of Mart\'\i n et al. (1999) and Kirkpatrick et al. (1999) agree for L2.}.

\section{Discussion}

HD~130948 is a chromospherically active single G2V star with high lithium abundance, and fast rotation (P=7.8 days). 
All these properties are indicative of youth (age$<$0.8~Gyr; Gaidos et al. 2000).  
The space motions of HD~130948 suggest that it could be related to the Ursa Major stream (Fuhrmann 2002, in preparation), which has an age of about 300~Myr. 

The two new companions of HD~130948 are probably contracting brown dwarfs because of the young age of the primary star. 
With the aim of estimating their ages and masses, we placed HD~130948~B and C  
on an H-R diagram with theoretical evolutionary tracks shown in Figure~3. We used the evolutionary models of Chabrier et al. (2000) that include 
dust in the equation of state and the opacity because those are appropiate for L dwarfs (Allard et al. 2001). 
Basri et al. (2000) and Schweitzer et al. (2001) 
have estimated the effective temperatures (T$_{\rm eff}$) of L dwarfs using the dusty models of  
Allard et al. (2001). 
Leggett et al. (2001) have used the same atmosphere models plus structural models for objects 
of known distance. 
We adopt T$_{\rm eff}$=1950$\pm$250~K for HD~130948B and C, which includes the whole range of T$_{\rm eff}$ estimates 
for L0--L4 dwarfs 
in the literature. Our NIRSPEC data alone are not sufficient to tell whether HD~130948B and C have different L spectral type because 
the region that we observed does not contain features that are sensitive to changes in subclass in the range L0 to L4. 
We note, however, that if we force HD~130948B and C to lie on the same isochrone, their spectral types 
should differ by about 2 subclasses. Further spectroscopic observations, particularly at optical wavelengths, 
can test the agreement between the position of these objects in the H-R diagram and the model predictions.   

For an age younger than 1~Gyr (consistent with youth of HD~130948A), 
the Chabrier et al. (2000) dusty models give a mass less than 0.075~M$_{\odot}$ for HD~130948B, 
and less than 0.065~M$_{\odot}$ for HD~130948C. It is very likely that both objects are young contracting brown dwarfs. 
For a total mass of the binary system of about 0.013~M$_{\odot}$, and a semimajor axis of 2.4~AU, the orbital period should be 
$\sim$10 years. Follow-up observations of this binary system over the next few years will yield dynamical masses for 
these two L dwarfs, which will extend the mass-luminosity-spectral type relation to cooler temperatures, and 
will provide two well constrained calibration points for the theoretical models describing low-mass, ultracool objects. 

Although there are a handful of brown dwarfs known as companions to main sequence stars, HD130948 B-C is the first brown dwarf binary system imaged around a G-type star. This advance has been rendered possible by the high dynamic range provided by the Hokupa'a AO system on the Gemini-North telescope. At the time of writing this paper, 31 G-type stars less than 1Gyr old have been observed with Hokupa'a/Gemini in our ongoing survey for VLM companions to the stars in the Gaidos et al. (2000) sample and other nearby, young G-type stars. The survey observations are sensitive to objects 2 magnitudes fainter than the HD130948B-C objects at radii inbetween 10 AU and 100 AU from the stars. The detection of this new binary brown dwarf system 
in our survey 
gives a 3.2\% $\pm$3.2\% frequency of brown dwarfs in the radius region we are sensitive to. 
This number is likely a lower limit because we are not sensitive to low mass brown dwarfs. 
Gizis et al. (2001) have reported a frequency of brown dwarf companions to G-type stars 
of 18\% $\pm$ 14\% for separations larger than 1000 AU. Liu et al. (2002) have found an L-type 
companion at 14~AU of a G-type star using adaptive optics. 
Combining our result with that of Gizis et al. and Liu et al., we 
suggest that brown dwarf companions to G-type dwarfs with separations larger than 10~AU 
may be common. The brown dwarf desert may be restricted to separations less than 10~AU. 
This supports the theoretical models of Armitage \& Bonnell (2002) that explain a lack 
of brown dwarfs within 10~AU of solar type stars as a consequence of orbital migration in 
circumstellar disks.


\acknowledgments
Support for proposal 9157 was provided by NASA through a grant 
from the Space Telescope Science Institute, which is operated 
by the Association of Universities for Research in Astronomy, Inc., 
under NASA contract NAS5-26555, as well for contract NAG-8298.

The Hokupa'a adaptive optics instrument is supported by NSF grant AST-9618852. This research has made use of the Simbad database, operated at CDS, Strasbourg, France. The authors would like to thank the Gemini North Observatory System Support Associates (SSA's), as well as Mark Chun, Kathy Roth, and Francios Rigaut for their professional support on the Gemini observing runs.
\clearpage

\clearpage

\figcaption[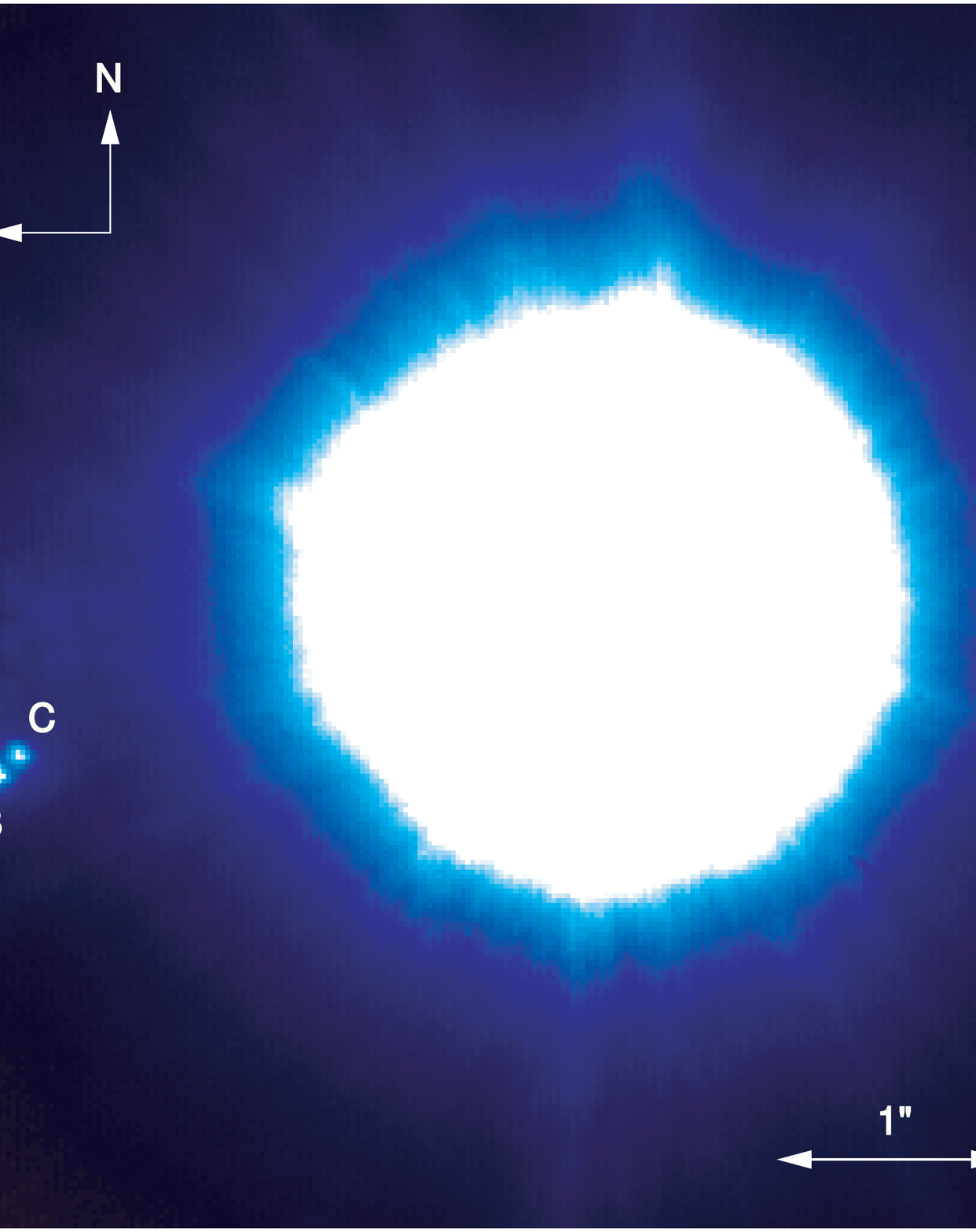]{The discovery image from the night of 2001 February 24 (UT) clearly shows the binary companion 
in individual 20 second exposures. The binary pair is 2$\farcs$64$\pm$0$\farcs$01 from and 8 H-band magnitudes fainter than HD130948A. The companion furthest from the 
primary is the brightest, thus we label it as HD130948B and the fainter companion as HD130948C. The separation between B and C on April 19 was 0$\farcs$134$\pm$0$\farcs$005 and the orientation PA=317$^{o}\pm$1$^{o}$.   \label{fig1}}

\figcaption[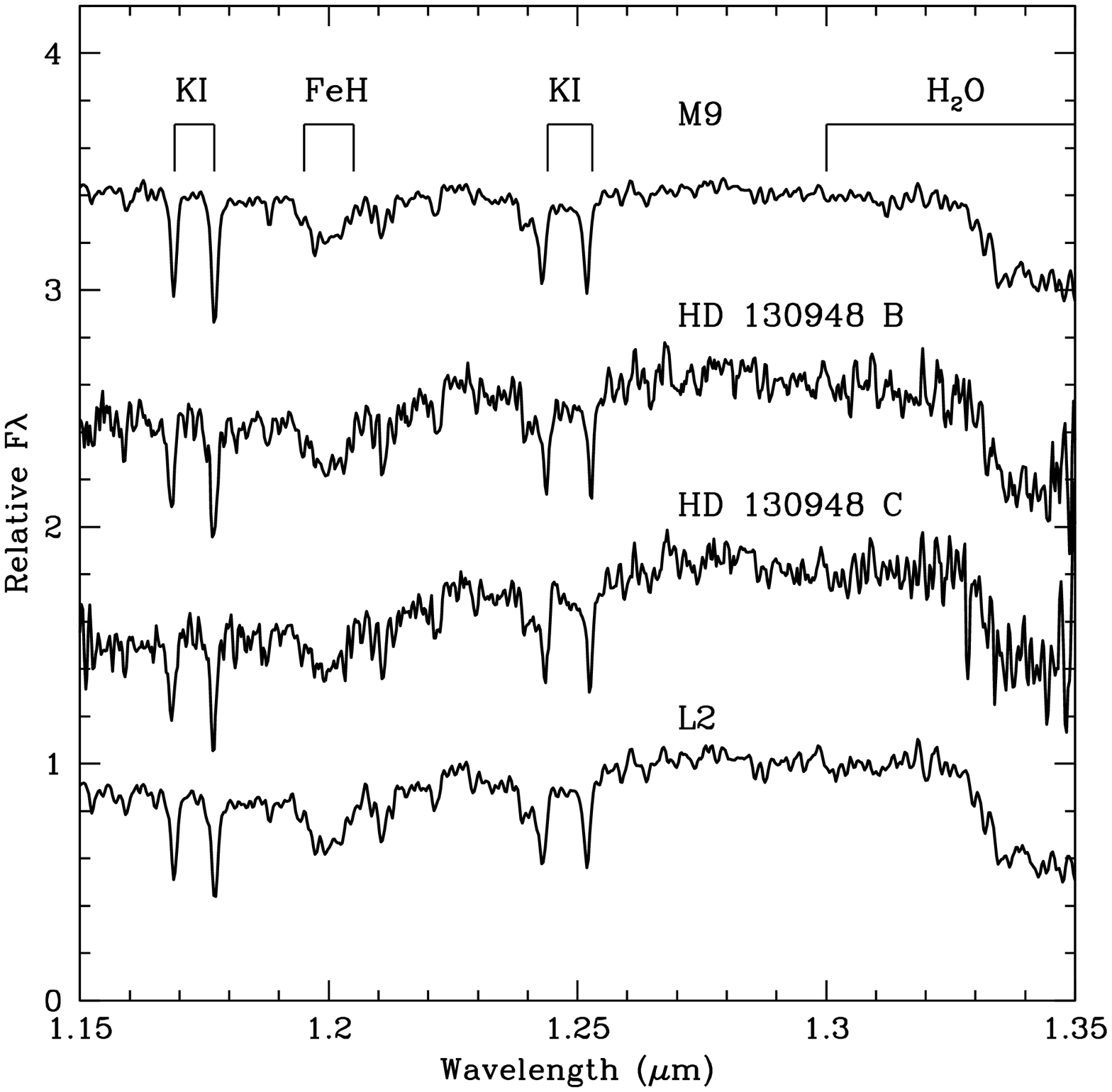]{Keck/NIRSPEC spectra of HD130948 B and C compared with IRTF/Spex spectra of known VLM objects. 
The M9 dwarf is DENIS-P J104814-395606 (Delfosse et al. 2001), and the L2 dwarf is Kelu~1 (Ruiz et al. 1997). 
\label{fig2}}

\figcaption[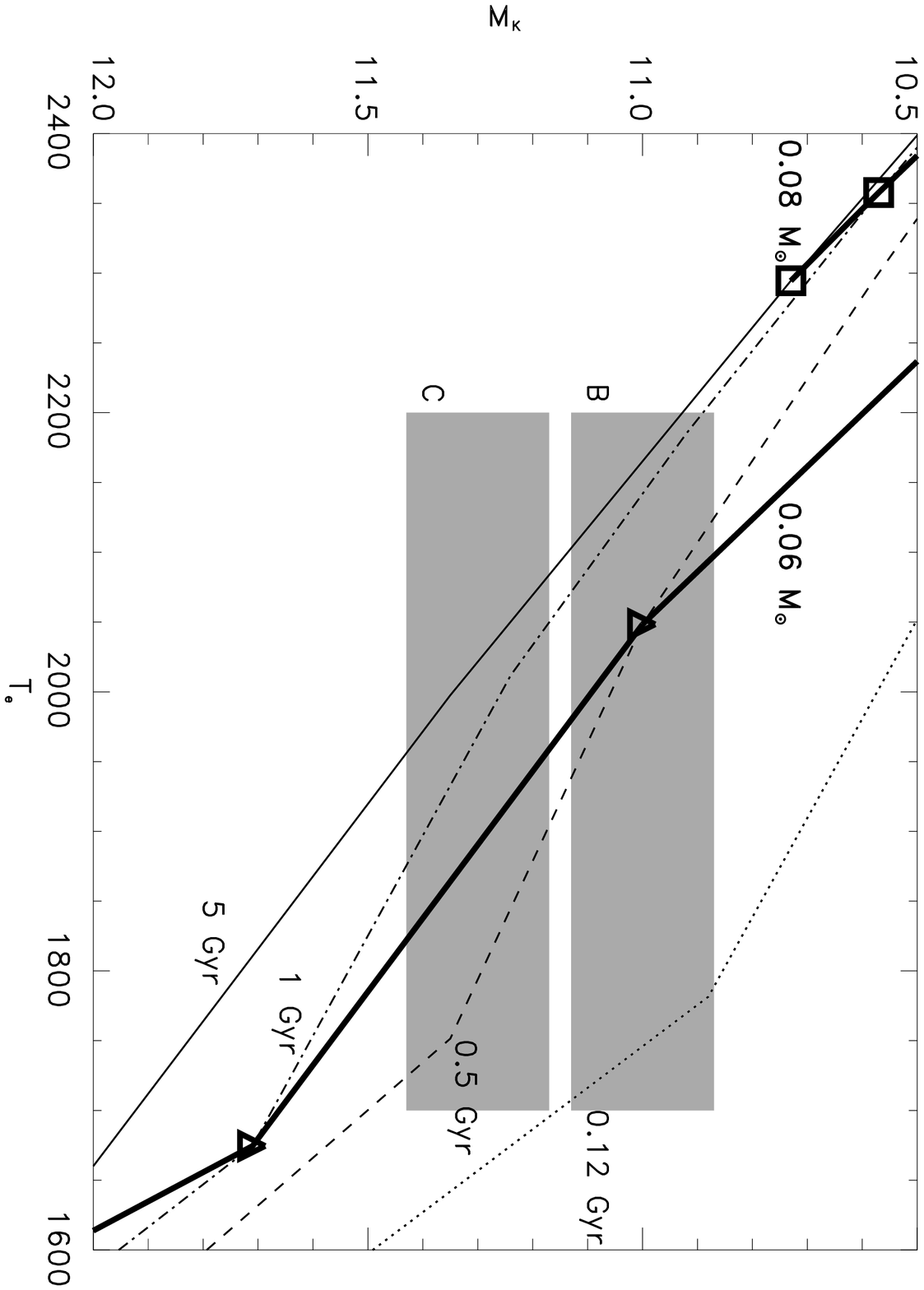]{The positions of the companions to HD130948 on the H-R diagram are compared to the dusty models of Chabrier et al. (2000). 
The 0.12, 0.5, 1.0, and 5.0 Gyr isocrones are distinguished by different line styles and are labeled.  The masses corresponding to the isocrones are plotted as bold squares and triangles connected with bold lines for 0.08 M$_{\odot}$ and 0.06 M$_{\odot}$ respectively. The shaded boxes labled B and C mark the range of temperature (x-axis) and photometric (y-axis) error values.  The range of T$_{\rm eff}$ corresponds to our measured range of spectral types (dL2$\pm$2). \label{fig3}}

\clearpage

\plotone{f1.eps}

\plotone{f2.eps}

\plotone{f3.eps}

\end{document}